# *gghic*: A Versatile R Package for Exploring and Visualizing 3D Genome Organization

Minghao Jiang[1], Duohui Jing[2] and Jason W.H. Wong[1,3,*]

[1]School of Biomedical Sciences, Li Ka Shing Faculty of Medicine, The University of Hong Kong, Hong Kong SAR, China

[2]Shanghai Institute of Hematology, State Key Laboratory of Medical Genomics, National Research Center for Translational Medicine at Shanghai, Ruijin Hospital Affiliated to Shanghai Jiao Tong University School of Medicine, Shanghai, China

[3]Centre for PanorOmic Sciences, The University of Hong Kong, Pokfulam, Hong Kong SAR, China

[*]Corresponding author

Email: jwhwong@hku.hk (J.W.H.W.)

Tel: +852 3917 9187

Fax: +852 2817 0857

**Abstract**

**Motivation:** The three-dimensional (3D) organization of the genome plays a critical role in regulating gene expression and maintaining cellular homeostasis. Disruptions in this spatial organization can result in abnormal chromatin interactions, contributing to the development of various diseases including cancer. Advances in chromosome conformation capture technologies, such as Hi-C, have enabled researchers to study genome architecture at high resolution. However, the efficient visualization and interpretation of these complex datasets remain a major challenge, particularly when integrating genomic annotations and inter-chromosomal interactions.

**Results:** We present *gghic*, an R package that extends the *ggplot2* framework to enable intuitive and customizable visualization of genomic interaction data. *gghic* introduces novel layers for generating triangular heatmaps of chromatin interactions and annotating them with features such as chromatin loops, topologically associated domains (TADs), gene/transcript models, and data tracks (*e.g.*, ChIP-seq signals). The package supports data from multiple chromosomes, facilitating the exploration of inter-chromosomal interactions. Built to integrate seamlessly with the R/Bioconductor ecosystem, *gghic* is compatible with widely used genomic data formats, including *HiCExperiment* and *GInteractions* objects. We demonstrate the utility of *gghic* by replicating a published figure showing a translocation event in T-cell acute lymphoblastic leukemia (T-ALL), highlighting its ability to integrate genomic annotations and generate publication-quality figures.

**Availability and implementation:** The R package can be accessed at https://github.com/jasonwong-lab/*gghic* and is distributed under the GNU General Public License version 3.0.

**Introduction**

DNA in mammalian cells is not distributed in a linear fashion inside the micrometer-size nucleus; instead, it is precisely organized under non-stochastic regulations into the hierarchical structure of higher-order chromatin fibers crucial for various activities, for example, cell division and homeostasis (Dekker, et al., 2017; Yu and Ren, 2017; Zheng and Xie, 2019). This spatial organization of the genome within the nucleus is referred to as the three-dimensional (3D) genome, which is dynamic in response to cellular signals and environmental factors. The 3D genome conformation is considered a vital aspect of epigenetic regulation, and any abnormalities within it can disrupt the precisely regulated active/inactive chromatin compartments (A/B compartments), topologically associated domains (TADs), and chromatin loops, potentially leading to the development of various diseases, including cancer (Akdemir, et al., 2020; Iyyanki, et al., 2021; Norton and Phillips-Cremins, 2017; Wang, et al., 2021; Watanabe, et al., 2024; Xu, et al., 2022; Xu, et al., 2022; Yang, et al., 2021).

To gain a deeper understanding of 3D genome organization, various techniques have been developed (Jerkovic and Cavalli, 2021; Wang, et al., 2021). Among them, the creation of chromosome conformation capture (3C) and its derivatives, *e.g.*, circular 3C (4C) (Zhao, et al., 2006), Hi-C (Belton, et al., 2012), Micro-C (Hsieh, et al., 2015), Pore-C (Deshpande, et al., 2022; Ulahannan, et al., 2019), and HiPore-C (Zhong, et al., 2023), has significantly advanced the study of nuclear architecture. By utilizing short-read and long-read sequencing technologies, researchers can capture pairwise and multi-way chromatin contacts across the entire genome, achieving resolutions as fine as 1 kilobase pair (Jerkovic and Cavalli, 2021). Briefly, DNA is sequenced after cell crosslinking, enzyme digestion, and DNA fragment ligation, producing sequencing reads containing DNA sequences physically proximal in the nucleus. Given the raw sequences, a matrix containing entries of chromatin contacts is generated by upstream bioinformatic pipelines for further downstream analysis, for example, the identification of

TADs and chromatin loops. Because of the large size and complexity of the sequencing data, efficient visualization and interpretation are required, with the UCSC Genome Browser (Perez, et al., 2024), Integrative Genomics Viewer (IGV) (Robinson, et al., 2011), and Juicebox (Durand, et al., 2016) being the most popular platforms for easy and intuitive usage. However, code-based packages/modules are more practical in terms of reproductivity and automation (Kramer, et al., 2022). Especially in the R and Bioconductor ecosystem (Gentleman, et al., 2004), many packages facilitating storing and manipulating genomic ranges and annotations and the *ggplot2* framework have become fundamental to high-throughput data analysis and visualization (Wickham, 2009).

While the R/Bioconductor community excels in data visualization, only a handful of R packages are adept at plotting genomic interaction data derived from 3D genome research (Ou and Zhu, 2019; Serizay, et al., 2024). This challenge becomes even greater when attempting to visualize genomic interactions across multiple chromosomes and presenting them in a triangular format to efficiently emphasize key features on the heatmap and beneath it. To fill this gap, we have developed an R package called *gghic*, which builds upon the widely used *ggplot2* package. *gghic* introduces new *Stat* and *Geom* objects, enabling users to create publication-ready triangular heatmaps of genomic interaction data, with other useful annotation layers for chromosome ideograms, gene/transcript annotations, data tracks, TADs, and chromatin loops. In addition, compatible with R/Bioconductor workflows, genomes, and annotations, *gghic* provides users seamless experience for Hi-C/-like data exploration.

**Materials and Methods**

*Overview of gghic*

*gghic* enhances *ggplot2* by introducing new *Stat* and *Geom* objects specifically designed for visualizing genomic interaction data. These objects enable the creation of triangular heatmaps, along with annotations for chromosome ideograms, gene/transcript models, data tracks, TADs, and chromatin loops. Users can input interaction data as a *data.frame/tibble* (Kirill and Hadley, 2023), *HiCExperiment* object (Serizay, et al., 2024), or *GInteractions* object (Lun, et al., 2016), making the package compatible with standard R/Bioconductor workflows (Figure 1A).

*Visualization of genomic interaction data*

In *gghic*, we developed a new layer called *geom_hic()* designed to generate triangular heatmaps for genomic interaction data. This layer requires the key aesthetics (seqname1, seqname2, start1, start2, end1, end2, and fill) to represent chromosome names, start and end positions, and the contact frequency for pairwise interactions. Notably, this layer can process interaction data from multiple chromosomes.

*Annotations above the heatmap*

A new layer, *geom_ideogram()*, was developed to produce chromosome ideograms above the triangular heatmap. In this layer, users can specify an assembly ID available from the UCSC Genome Browser. Regions displayed in the heatmap will be highlighted in the ideogram(s). Cytoband information, retrieved using the R package *rtracklayer* (Lawrence, et al., 2009), is cached for faster execution when the same assembly ID is requested again.

*Annotations under the heatmap*

Gene and transcript annotation tracks can be added below the triangular heatmap after calling *geom_annotation()*. To retrieve annotation information, users should provide either a *TxDb* object or a GTF file path. Users can choose to plot either specific genes/transcripts of interest or all genes/transcripts within the regions represented by the triangular heatmap. The generated *TxDb* and tx2gene *tibble* will be cached for efficiency.

Additionally, *geom_track()* was developed to help users add other genomic data tracks beneath the main heatmap. The maximal ranges for different tracks can be set to the maximum of each individual track, the maximum across all tracks, or can be manually specified with given values.

*Annotations on the heatmap*

The functions *geom_tad()* and *geom_loop()* were developed to enhance heatmaps by adding lines that represent TADs or compartments and by adding circles to indicate chromatin loops, respectively.

*A wrapper function for easy visualization of genomic interaction data and decoration*

Users can use a single function *gghic()* to generate a publication-ready figure for their genomic interaction data. This function consolidates all the aforementioned layers and slightly extends the x-axis to accommodate any text that might fall out of bounds. Moreover, the wrapper function utilizes a standard color theme frequently employed by various tools for generating Hi-C heatmaps. Since *gghic* is built upon *ggplot2*, it seamlessly integrates all *ggplot2*'s functionalities, allowing users to easily customize the color scheme and other details according to their preferences.

**Results**

The *gghic* package, designed to generate genomic interaction figures, requires input in the form of either a *data.frame/tibble* object or a *HiCExperiment* object (Figure 1A). As an extension of *ggplot2*, *gghic* introduces new layers that enable users to apply a specific set of aesthetic mappings defined using the *aes()* function within *ggplot()* (Materials and Methods). A total of six additional layers were created to draw the main triangular heatmap, chromosome ideograms above the heatmap, gene/transcript annotations under the heatmap, other data tracks under the heatmap, and TADs and loops on the heatmap (Figure 1A). The novel layer designs ensure that users can directly apply the grammar of graphics to a table of genomic interactions and control details using other functions from *ggplot2*, reducing the learning curve for users to become familiar with the package.

To demonstrate the utility of *gghic*, we applied it to publicly available Hi-C and ChIP-seq datasets (Cuddapah, et al., 2009; Hnisz, et al., 2016; Yang, et al., 2021) to replicate a published figure showing a translocation event in T-cell acute lymphoblastic leukemia (T-ALL) (Figure 1B) (Yang, et al., 2021). The figure highlights a *BCL11B::TLX3* translocation, with chromatin loops and TADs annotated on the heatmap. ChIP-seq tracks for CTCF and H3K27ac signals were added below the heatmap to validate the hijacked enhancer activity. Ideograms of the translocated chromosomes were included above the heatmap. The resulting figure reproduced the published visualization with high fidelity, demonstrating *gghic*'s ability to generate publication-ready figures while handling data from multiple chromosomes and integrating diverse genomic annotations (Figure 1B) using the standard R code (Figure 1C).

To effectively manage large datasets, *gghic* allows for the creation of rasterized heatmaps while maintaining other components as vectors. We tested the package with a genomic interaction dataset containing 34,737,001 pairwise contacts across the entire human genome at a 50 kilobase resolution. Plotting this dataset took approximately 4.62 minutes on a Linux server with 754GB of memory. In comparison, the R package *HiContacts* took about 8.36 minutes using the *plotMatrix* function, while the Python package *Matplotlib* took around 1.13 minutes using the *matplotlib.pyplot.matshow* function. Although *gghic* is not the fastest tool compared to some optimized Python implementations, it balances performance with extensive flexibility and integration capabilities within the R environment. Notably, researchers often focus on visualizing smaller subregions with detailed annotations, an area where *gghic* truly excels. Its ability to provide detailed genomic annotations and fully customizable visual outputs is crucial. Furthermore, *gghic*'s seamless integration with other R/Bioconductor packages allows for a comprehensive analytical workflow, from data processing to visualization, within a single environment.

**Discussion and Conclusion**

The *gghic* R package was developed to address the growing need for an intuitive and flexible tool to visualize genomic interaction data derived from Hi-C and related technologies. By leveraging the powerful *ggplot2* framework, *gghic* integrates seamlessly into the R/Bioconductor ecosystem, offering a user-friendly yet highly customizable solution for generating publication-ready figures. The package introduces novel *Stat* and *Geom* objects to facilitate the visualization of triangular heatmaps, chromatin loops, TADs, and other genomic annotations, filling a critical gap in the existing R packages available for genomic interaction analysis.

One of the key strengths of *gghic* lies in its ability to adapt to the complexity of genomic interaction datasets. The package supports data from multiple chromosomes, allowing researchers to explore inter-chromosomal interactions, which are of particular interest in cancer genomics but often overlooked by existing visualization tools. Furthermore, *gghic* provides an efficient workflow for integrating additional genomic annotations such as ideograms, gene/transcript models, and data tracks (*e.g.*, ChIP-seq signals), enabling a more comprehensive interpretation of 3D genome organization and its functional implications. The package's compatibility with widely used formats in the R/Bioconductor ecosystem, such as *HiCExperiment* and *GInteractions* objects, ensures that users can seamlessly incorporate *gghic* into existing analysis pipelines.

In our demonstration, we successfully applied *gghic* to visualize a translocation event in a T-cell acute lymphoblastic leukemia (T-ALL) case, replicating a figure from a previous publication (Yang, et al., 2021). The ability to include chromatin loops, TADs, and ChIP-seq tracks in a single visualization highlights the utility of *gghic* for understanding the interplay between 3D genome architecture and transcriptional regulation. The package's flexibility in customizing visual elements, such as color schemes and annotation styles, ensures that researchers can tailor the output to meet the specific needs of their studies.

Despite its strengths, *gghic* has certain limitations. The package's reliance on *ggplot2* can lead to performance challenges when handling very large datasets, particularly at high resolutions. Future development will focus on optimizing the computational efficiency of the package and exploring compatibility with other visualization frameworks. Additionally, while *gghic* provides extensive support for genomic annotations, further integration with external databases (*e.g.*, ENCODE, Roadmap Epigenomics) could enhance its utility for functional genomics studies (Bernstein, et al., 2010; Kagda, et al., 2023).

In conclusion, *gghic* is a powerful and versatile R package for visualizing genomic interaction data, offering an intuitive interface for creating complex, publication-quality figures. By bridging the gap between genomic data visualization and the grammar of graphics, *gghic* empowers researchers to explore and communicate their findings on 3D genome organization more effectively. We anticipate that *gghic* will become a valuable resource for the genomics community, facilitating deeper insights into spatial dynamics of genomes and their implications for health and disease.

**Data Availability**

All data used in this study are publicly accessible. Processed datasets, including the Wig file for H3K27ac ChIP-seq of Jurkat cells, the Hi-C matrix in hic format of the T-ALL patient, and the BigWig file for H3K27ac ChIP-seq of the T-ALL patient, were obtained from the Gene Expression Omnibus (GEO) under accession numbers GSE68978 and GSE146901. Raw data for CTCF ChIP-seq of Jurkat cells were retrieved from the Sequence Read Archive (SRA) under accession number SRR014986.

**Funding**

This research was funded by grants from the National Key R&D Program of China (2022YFE0200100, 2023YFC2508901) and the Mainland-Hong Kong Joint Funding Scheme supported by the Innovation and Technology Commission, the Government of Hong Kong SAR, China (MHP/054/21).

**Conflict of Interest**

None declared.

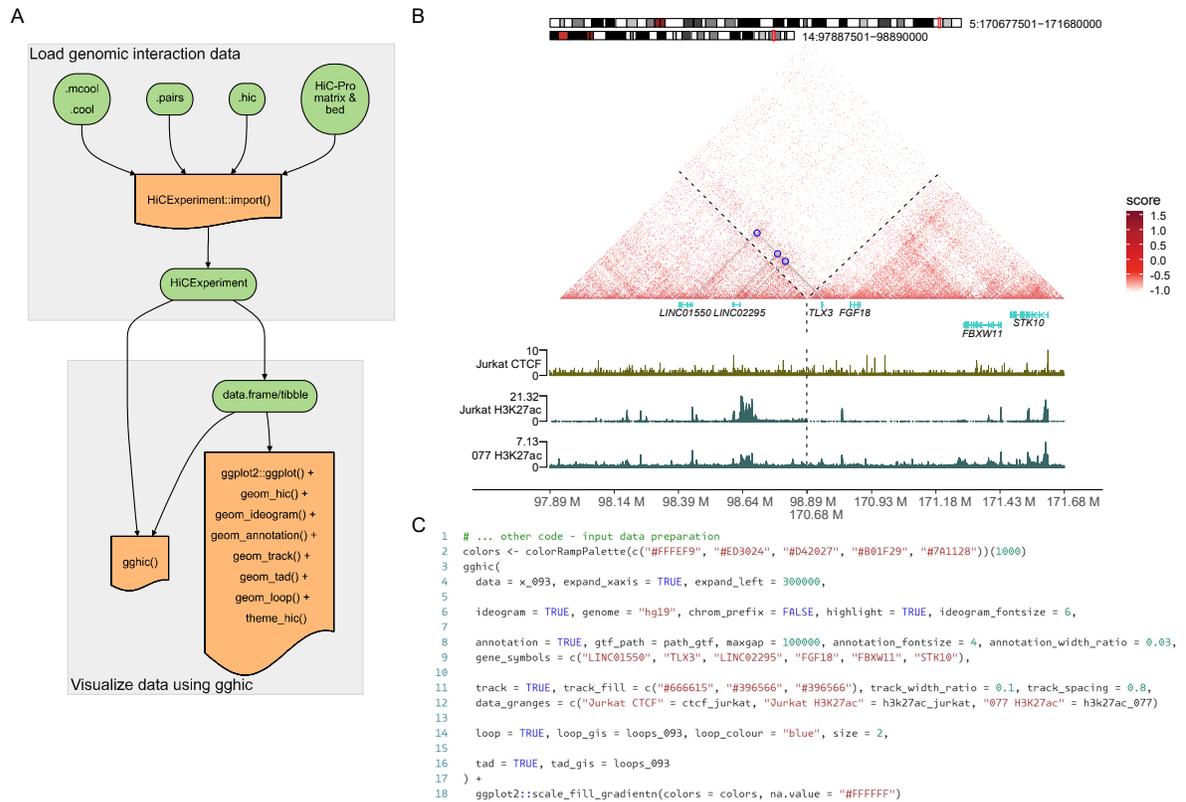

**Figure 1. A simple guide and example for using *gghic*.** (A) A flowchart illustrating the workflow for using the *gghic* package. (B) A heatmap showcasing a T-cell acute lymphoblastic leukemia (T-ALL) case with a *BCL11B::TLX3* translocation. The heatmap is accompanied by ChIP-seq tracks for CTCF and H3K27ac from Jurkat cells, as well as an H3K27ac ChIP-seq track from another T-ALL case (sample ID: 077). Highlighted chromatin loops are marked with blue circles and gray dashed lines. This serves as an example of reproducing Figure 3e from Yang *et al.*'s publication using the *gghic* package. (C) The source code for generating the figure.